\documentclass{jpsj3}
\usepackage{txfonts}

\title{Observation of Low-Energy Einstein Phonon and Superconductivity in Single-Crystalline LaBe$_{13}$}

\author{Hiroyuki Hidaka$^1$\thanks{hidaka@phys.sci.hokudai.ac.jp}, Yusei Shimizu$^1$\thanks{Present address: Institute for Materials Research, Tohoku University, Oarai, Ibaraki 311-1313, Japan}, Seigo Yamazaki$^1$, Naoyuki Miura$^1$, Ryoma Nagata$^1$, Chihiro Tabata$^1$\thanks{Present address: Condensed Matter Research Center and Photon Factory, Institute of Materials Structure Science, High Energy Accelerator Research Organization, Tsukuba, Ibaraki 305-0801, Japan}, Shota Mombetsu$^1$, Tatsuya Yanagisawa$^1$, and Hiroshi Amitsuka$^1$}

\inst{$^1$Graduate School of Science, Hokkaido University, Sapporo, Hokkaido 060-0810, Japan \\
} %\\

\abst{The thermal and electrical transport properties of single-crystalline LaBe$_{13}$ have been investigated by specific-heat ($C$) and electrical-resistivity ($\rho$) measurements.
The specific-heat measurements in a wide temperature range revealed the presence of a hump anomaly near 40 K in the $C$($T$)/$T$ curve, indicating that LaBe$_{13}$ has a low-energy Einstein-like-phonon mode with a characteristic temperature of $\sim$ 177 K. 
In addition, a superconducting transition was observed in the $\rho$ measurements at the transition temperature of 0.53 K, which is higher than the value of 0.27 K reported previously by Bonville $et$ $al$. 
Furthermore, an unusual $T^3$ dependence was found in $\rho$($T$) below $\sim$ 50 K, in contrast to the behavior expected from the electron--electron scattering or the electron--Debye phonon scattering. 
}

\begin{document}
\maketitle

%\section{Introduction}

The beryllides MBe$_{13}$ (M = rare earths and actinides) show several novel physical properties depending on the M atom, such as unconventional superconductivity (SC) in UBe$_{13}$ \cite{Ott}, an intermediate valence state in CeBe$_{13}$\cite{Cooper}, and helical-magnetic ordering in HoBe$_{13}$ \cite{Vigneron}. 
The MBe$_{13}$ compounds crystallize in an isostructural NaZn$_{13}$-type cubic structure with the space group $F$$m$$\bar{\rm 3}$$c$ (No. 226, $O_h^{\rm 6}$). 
The unit cell contains eight formula units, where the M atoms occupy eight equivalent positions in the 8$a$ site with the cubic site symmetry $O$, while the Be atoms occupy two different crystallographic sites: Be$^{\rm I}$ (8$b$) and Be$^{\rm II}$ (96$i$) \cite{Mcelfresh, Takegahara}. 
A significant feature of this structure is that the M ions are surrounded by 24 Be$^{\rm II}$ atoms, nearly forming a snub cube, whereas the Be$^{\rm I}$ atoms are surrounded by 12 Be$^{\rm II}$ atoms, forming an icosahedron, namely, the unit cell consists of two cagelike structures.

Recently, such cage-structured systems, as represented by filled skutterudites and $\beta$-pyrochlores, have been attracting much attention because of the presence of a low-energy phonon mode associated with local vibration of a guest atom with a large amplitude in an oversized host cage, called rattling \cite{Ogita, Iwasa, Matsuhira, Hiroi}. 
The presence of rattling is considered to be related to some novel phenomena, such as the magnetic-field-robust heavy-fermion state in SmOs$_4$Sb$_{12}$ \cite{Yana, Hattori} and the strong-coupling $s$-wave SC in KOs$_2$O$_6$ \cite{Hiroi}. 
In the MBe$_{13}$ systems, previous inelastic-neutron-scattering (INS) measurements revealed that UBe$_{13}$ and ThBe$_{13}$ have a low-energy Einstein-like-phonon mode with an Einstein temperature $\theta_{\rm E}$ of $\sim$ 151 and 157 K, respectively \cite{Renker}. 
A systematic study of the low-energy Einstein-like-phonon mode in the MBe$_{13}$ series would be useful for understanding the anomalous physical properties in UBe$_{13}$ \cite{Ott}, such as unconventional SC and non-Fermi-liquid behavior. 
However, it is unclear whether other MBe$_{13}$ compounds also have a low-energy Einstein-like-phonon mode.

In the present study, we focus our attention on LaBe$_{13}$, since the study on a La-based compound without 4$f$ electrons should be essential to understand physical properties in other MBe$_{13}$ systems with $f$ electrons. 
Specific-heat ($C$) measurements of LaBe$_{13}$ were performed and reported by three groups more than thirty years ago, where the authors estimated the electronic specific-heat coefficient $\gamma$ and the Debye temperature $\theta_{\rm D}$ to be as follows: $\gamma$ = 0.58 mJmol$^{-1}$K$^{-2}$ and $\theta_{\rm D}$ = 820 K by Bucher $et$ $al$. \cite{Bucher}; $\gamma$ =11 mJmol$^{-1}$K$^{-2}$ and $\theta_{\rm D}$ = 554 K by Van Der Linden $et$ $al$. \cite{Linden}; $\gamma$ = 7.8 mJmol$^{-1}$K$^{-2}$ and $\theta_{\rm D}$ = 865 K by Besnus $et$ $al$. \cite{Besnus}. 
However, these temperature ranges in the previous studies were limited to below 40 K, and no comments about the Einstein mode were given. 
From electrical-resistivity ($\rho$) measurements, Bucher $et$ $al$. mentioned that a superconducting transition is absent above 0.45 K \cite{Bucher}. 
After that, Bonville $et$ $al$. found SC with $T_{\rm SC}$ $\sim$ 0.27 K by the M$\ddot{\rm {o}}$ssbauer, $\rho$ and dc magnetic-susceptibility measurements \cite{Bonville}. 
However, no experimental details and data concerning the SC are described in their paper. 
It should be noted that all the previous studies were performed by using polycrystalline samples. 
The present paper is the first report on the thermal and electrical transport properties of single-crystalline LaBe$_{13}$, and we found evidence of the presence of the low-energy Einstein-like phonon as well as SC with $T_{\rm SC}$ $\sim$ 0.53 K, higher than that obtained for the polycrystalline sample.

%\section{Experimental Details}

Single crystals of LaBe$_{13}$ were grown by the Al-flux method. 
The constituent materials (La with 99.9$\%$ purity and Be with 99.9$\%$ purity) and Al with 99.99$\%$ purity were placed in an Al$_{2}$O$_{3}$ crucible at an atomic ratio of 1:13:35 and sealed in a quartz tube filled with ultrahigh-purity Ar gas of 150 mmHg. 
The sealed tube was kept at 1050 $^\circ$C for 72 h and then cooled at a rate of 2 $^\circ$C/h. 
The Al-flux was spun off in a centrifuge and then removed by NaOH solution. 
The prepared samples were annealed for 2 weeks at 700 $^\circ$C. 
After sample annealing, some impurity appeared to have been educed on the surface.

The grown samples were characterized by powder X-ray diffraction (XRD) using Cu-K$\alpha_{1}$ and K$\alpha_{2}$ radiation as shown in Fig. 1. 
The obtained XRD pattern of the annealed LaBe$_{13}$ does not show any impurity phases within the experimental accuracy except for reflections come from a copper holder. 
The lattice parameter was determined as $a$ = 10.506(3) $\rm {\AA}$ from the high-angle data (2$\theta$ $>$ $110^\circ$), which is slightly larger than the reported value of 10.451 $\rm {\AA}$ \cite{Bucher}. 
In addition, we performed Rietveld analysis based on the NaZn$_{13}$-type cubic structure, where the lattice parameter was fixed to 10.506 $\rm {\AA}$. 
The typical reliability factors are $R_{\rm wp}$ = 7.76$\%$, $R_{\rm F}$ = 2.69$\%$, and $S$ = 1.71. 
From the present analysis, fractional (0, $y$, $z$) coordinates of the Be$^{\rm II}$ site were obtained as $y$ = 0.17607(147) and $z$ = 0.11373(147). 
These values are close to previously reported values for UBe$_{13}$ ($y$ = 0.1763(1), $z$ = 0.1150(1)) \cite{Goldman} and for an ideal snub cube ($y$ = 0.17610, $z$ = 0.11408) \cite{Hudson}.

\begin{figure}[tb]
\begin{centering}
\includegraphics[width=0.5\textwidth]{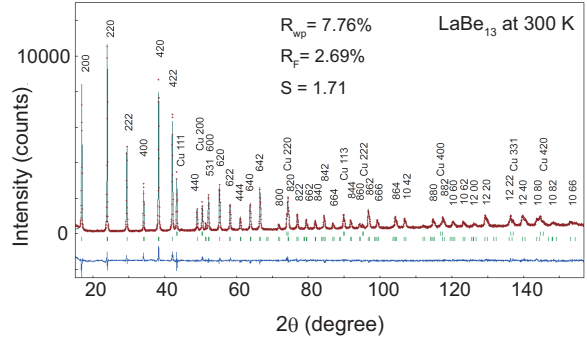}
\caption{
(Color online) Powder XRD pattern of LaBe$_{13}$ observed at 300 K (cross symbols), and calculated profile (solid line). 
The green vertical bars below the XRD pattern indicate the calculated peak positions. 
The difference between the observed and calculated intensities is indicated by the blue line at the bottom of the figure. 
}
\label{Fig1}
\end{centering}
\end{figure}

The specific-heat measurements were performed by a thermal-relaxation method in the temperature range of 2 -- 300 K with a commercial Physical Property Measurement System (PPMS, Quantum Design, Inc.). 
Electrical-resistivity measurements were performed by a conventional four-probe method in the temperature range of 0.3 K $\leq$ $T$ $\leq$ 300 K. 
In the present $\rho$ measurements, the current $\boldmath J$ was applied along the [100] direction.

%\section{Results and Discussion}

Figure 2 shows the temperature dependence of the specific heat divided by the temperature $C$($T$)/$T$ of LaBe$_{13}$ measured on an unannealed sample. 
In the low-temperature region below $\sim$ 10 K (see the inset of Fig. 2), the $C$($T$)/$T$ curve obeys the Debye $T^3$ law: $C$/$T$ = $\gamma$ + $\beta$$T^2$. 
The Debye temperature $\theta_{\rm D}$ can be determined from the following expression: 
\begin{equation} 
\theta_{\rm D} = (12\pi^{4}Rn/5\beta)^{1/3}, 
\end{equation} 
where $R$ is the gas constant and $n$ (= 14) is the number of atoms in the formula unit. 
From the experimental results, we determined $\gamma$ and $\theta_{\rm D}$ as 9.1 mJmol$^{-1}$K$^{-2}$ and $\sim$ 950 K, respectively \cite{Shimizu}. 
The obtained $\gamma$ is in good agreement with the value determined from the two previous experiments \cite{Linden, Besnus} and a band calculation \cite{Takegahara}, whereas the obtained $\theta_{\rm D}$ is somewhat higher than that reported in all the previous experimental studies \cite{Bucher, Linden, Besnus}.

\begin{figure}[tb]
\begin{centering}
\includegraphics[width=0.4\textwidth]{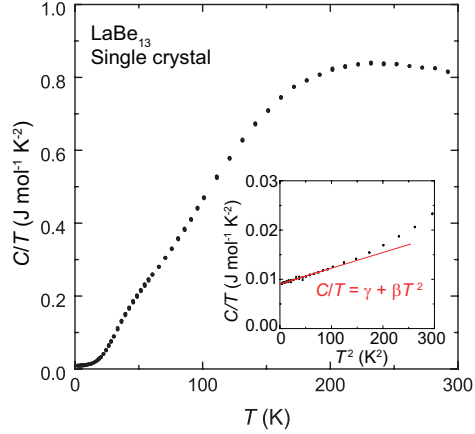}
\caption{(Color online) Temperature dependence of specific heat divided by temperature $C$/$T$ for single-crystalline LaBe$_{13}$ (unannealed). The inset shows the low-temperature part of $C$($T$)/$T$ as a function of $T^2$. The red line represents the Debye $T^3$ law. 
}
\label{Fig1}
\end{centering}
\end{figure}

\begin{figure}[tb]
\begin{centering}
\includegraphics[width=0.42\textwidth]{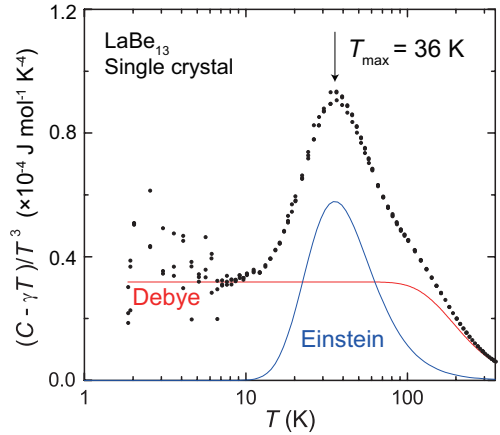}
\caption{(Color online) Temperature dependence of ($C$ -- $\gamma$$T$)/$T^3$ for LaBe$_{13}$. The red and blue lines represent the Debye and Einstein specific-heat contributions calculated using the obtained $\gamma$, $\theta_{\rm D}$, and $\theta_{\rm E}$.}
\label{Fig2}
\end{centering}
\end{figure}

The $C$($T$)/$T$ curve of LaBe$_{13}$ shows a broad hump anomaly near 40 K, which cannot be described only by the Debye phonon contribution. 
In Fig. 3, we plotted the temperature dependence of ($C$ -- $\gamma$$T$)/$T^3$ for LaBe$_{13}$, since such analysis is useful in investigating the Einstein phonon contribution of the specific heat. 
In the intermediate temperature range between $\sim$ 10 and $\sim$ 200 K, the ($C$ -- $\gamma$$T$)/$T^3$ curve considerably deviates from the Debye specific heat (the red solid line) and shows a broad peak at $T_{\rm max}$ $\sim$ 36 K. 
$T_{\rm max}$ is known to be linked to $\theta_{\rm E}$ by the relationship $\theta_{\rm E}$ $\sim$ 4.92 $T_{\rm max}$ \cite{Matsuhira}. 
From this relationship, we estimated the value of $\theta_{\rm E}$ for LaBe$_{13}$ to be $\sim$ 177 K\cite{Shimizu}, which is higher than that reported for La-based filled skutterudite compounds and $\beta$-pyrochlore oxides, for example, $\theta_{\rm E}$ = 138 K for LaFe$_4$P$_{12}$ \cite{Matsuhira} and $\theta_{\rm E}$ = 70 K for CsOs$_2$O$_6$ \cite{Hiroi2}. 
The Einstein contribution to the specific heat calculated from the obtained $\theta_{\rm E}$ (the blue solid line in Fig. 3) well reproduces the hump structure in the specific heat.

Note that such a hump structure in $C$($T$)/$T$ has been found near 40 K even for ThBe$_{13}$ \cite{Felten}, suggesting that $\theta_{\rm E}$ for ThBe$_{13}$ is close to that for LaBe$_{13}$. 
In addition, $\theta_{\rm E}$ of UBe$_{13}$ and ThBe$_{13}$ were estimated from the previous INS measurements to be $\sim$ 151 and 157 K, respectively \cite{Renker}. 
These findings suggest that the MBe$_{13}$ systems commonly possess a low-energy Einstein-like-phonon mode. 
If the low-energy phonon mode is attributed to the local oscillation of the guest atom in the cagelike structure, the difference in $\theta_{\rm E}$ between LaBe$_{13}$ and UBe$_{13}$/ThBe$_{13}$ might be explained by the difference in the mass between the M ions.

Figure 4 displays the temperature dependence of the electrical resistivity $\rho$($T$) of LaBe$_{13}$, where the measured sample was polished after annealing to remove the impurity on the surface. 
The present $\rho$($T$) measurements revealed that the annealing hardly affected both the temperature dependence and residual resistivity ratio (RRR); RRR = 3.5 and 3.3 for the unannealed and annealed samples, respectively. 
The $\rho$($T$) curve of LaBe$_{13}$ exhibits $T$-linear dependence above $\sim$ 150 K, showing simple metallic behavior. 
Dahm and Ueda have calculated $\rho$($T$) for an anharmonic rattling system, taking into account the coupling between conduction electrons and a local anharmonic phonon mode, and they predicted that $\rho$($T$) has $\sqrt{T}$ behavior at high temperatures \cite{Dahm}. 
Such $\sqrt{T}$ behavior has been observed in actual cage-structured systems with an anharmonic phonon mode, such as LaOs$_4$Sb$_{12}$\cite{Sugawara} and KOs$_2$O$_6$\cite{Hiroi2}, whereas it is not observed in LaBe$_{13}$. 
This suggests that the low-energy phonon mode of LaBe$_{13}$ may not be anharmonic.

\begin{figure}[tb]
\begin{centering}
\includegraphics[width=0.4\textwidth]{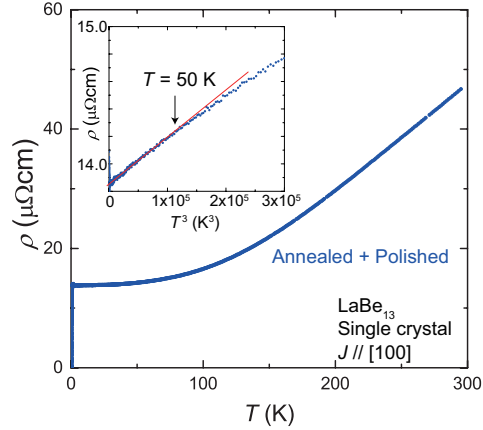}
\caption{(Color online) Temperature dependence of the electrical resistivity of LaBe$_{13}$ measured on the polished sample after annealing. The inset shows the $\rho$($T$) data at low temperatures as a function of $T^3$.}
\label{Fig3}
\end{centering}
\end{figure}

Instead, $\rho$($T$) for LaBe$_{13}$ shows $T^3$ behavior below 50 K, as shown in the inset of Fig. 4. 
This temperature dependence at low temperatures deviates from the expected behavior of simple metals: the electron--phonon scattering yields a $T^5$ dependence at low temperatures within the well-known Bloch--Gr$\ddot{\rm u}$neisen theory, assuming the Debye phonon and a spherical Fermi surface, whereas the electron-electron scattering gives a $T^2$ dependence in $\rho$($T$). 
In the case of LaBe$_{13}$ without 4$f$ electrons, one can exclude a contribution of the crystalline electric field or the magnetic correlations. 
One possible origin of the observed $T^3$ dependence is phonon-induced interband scattering between $s$ and $d$ bands, usually called $s$--$d$ interband scattering \cite{Webb, Colquitt}. 
It is known that the $T^3$ dependence due to the $s$--$d$ interband scattering can be seen in nonmagnetic transition metals with an incomplete $d$ shell. 
On the other hand, LaBe$_{13}$ does not contain any transition metals, making it doubtful that the $s$--$d$ interband scattering dominates the scattering mechanism of the present system. 
Note that in the band calculation for LaBe$_{13}$, the partial density of states of the La $d$ band accounts for only about 15$\%$ of the total density of states \cite{Takegahara}. 
Another possibility is that the low-energy Einstein-like phonon mode plays some role in the electron--phonon scattering, since the $T^3$ behavior in $\rho$($T$) is observed in the temperature range where the $C$($T$)/$T$ curve shows the hump anomaly. 
At the present stage, some questions about the low-energy Einstein-like-phonon mode in the MBe$_{13}$ systems still remain; which atom behaves as an Einstein oscillator and how large is its amplitude? 
To obtain further information, further studies, such as precise XRD and Raman scattering, will be needed.

\begin{figure}[tb]
\begin{centering}
\includegraphics[width=0.4\textwidth]{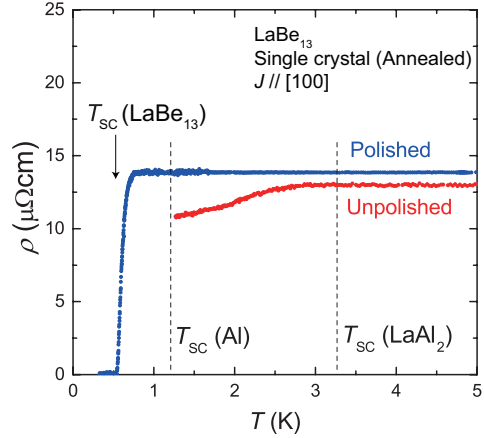}
\caption{(Color online) Electrical resistivity of LaBe$_{13}$ at low temperatures for the unpolished (red) and polished (blue) samples after annealing. $T_{\rm SC}$ for LaAl$_2$ and Al are also displayed in this figure for comparison.} 
\label{Fig4}
\end{centering}
\end{figure}

Figure 5 shows the low-temperature electrical resistivity of LaBe$_{13}$ below 5 K. 
For the unpolished sample after annealing (red data), the $\rho$($T$) curve shows a broad kink anomaly at $\sim$ 2.5 K. 
However, by polishing the sample surface, the kink anomaly disappears (see blue data), indicating that some impurity was educed on the surface by the annealing. 
The impurity is considered to be LaAl$_2$, since the temperature where the kink anomaly appears is close to $T_{\rm SC}$ for LaAl$_2$ (= 3.23 K) \cite{Smith}. 
For the polished sample, one can find a clear superconducting transition at $T_{\rm SC}$ = 0.53 K, which is different from those of Al (= 1.2 K), LaAl$_2$ (= 3.23 K), and LaAl$_{3}$ (= 5.57 K) \cite{Smith}, as indicated by dashed lines in Fig. 5. 
Here, $T_{\rm SC}$ is defined as the temperature where zero resistivity is achieved. 
Note that the SC of the present compound was observed even in the unannealed sample at the same $T_{\rm SC}$ (not shown). 
The $T_{\rm SC}$ observed in the present study is slightly higher than that reported previously by Bonville $et$ $al$. ($T_{\rm SC}$ = 0.27 K) \cite{Bonville} and is inconsistent with the previous result of Bucher $et$ $al$., where the SC was absent above 0.45 K \cite{Bucher}. 
This discrepancy might be due to a difference between the single crystal and the polycrystal and/or a difference in the purity of the constituent materials. 
Alternatively, $T_{\rm SC}$ for the polycrystal in Ref. 13) may have been lower than 0.45 K. 
To our knowledge, LaBe$_{13}$ is the second case of a superconductor after UBe$_{13}$ among the known MBe$_{13}$ systems. 
The superconducting properties of LaBe$_{13}$ are not yet well known except for $T_{\rm SC}$ at zero field. 
Very recently, specific-heat measurements have been performed down to $\sim$ 0.1 K, and thermodynamic evidence for the SC of the bulk has been found \cite{Shimizu2}.

%\section{Summary}

In summary, we have succeeded in growing single-crystalline LaBe$_{13}$ by the Al-flux method, and we performed $C$($T$) and $\rho$($T$) measurements. 
The $C$($T$)/$T$ curve shows a hump structure near 40 K, which can be interpreted as the contribution of the Einstein phonon with $\theta_{\rm E}$ $\sim$ 177 K. 
The obtained $\theta_{\rm E}$ is close to that of ThBe$_{13}$ and UBe$_{13}$ reported previously, suggesting that the low-energy Einstein-like phonon is commonly present in the MBe$_{13}$ systems. 
In the $\rho$($T$) measurements, we found an unusual $T^3$ dependence below $\sim$ 50 K, where the low-energy Einstein-like phonon might play some role in the electron-phonon scattering. 
In addition, a superconducting transition occurs at $T_{\rm SC}$ of $\sim$ 0.53 K, which is higher than the value in the previous report using a polycrystalline sample ($T_{\rm SC}$ = 0.27 K). 
Further studies will be required in order to reveal the detailed superconducting properties of LaBe$_{13}$.

\begin{acknowledgment}
%\section{Acknowledgments}

The present research was supported by JSPS Grants-in-Aid for Scientific Research (KAKENHI) Grant Nos. 20224015 (S), 25400346 (C), 26400342 (C), 15H05882, and 15H05885 (J-Physics).

\end{acknowledgment}

\end{document}